\documentclass[final]
  {aipproc}

\layoutstyle{6x9}

\usepackage{units}
\usepackage{amsmath}

\newcommand{\Gfermi}{\ensuremath{G_\mathrm{F}}}
\newcommand{\piEthree}{\ensuremath{\pi\mathrm{E3}}}

\begin{document}

\title{MuLan: Towards a \unit[1]{ppm} muon lifetime measurement}


\author{Kevin R. Lynch}{
  address={on behalf of the MuLan Collaboration\\Boston University
  Physics Department, 590 Commonwealth Ave, Boston, MA, 01824, USA},
  email={krlynch@bu.edu}
}

\begin{abstract}
The MuLan experiment will measure the lifetime of the positive muon to
\unit[1]{ppm}.  Within the Standard Model framework, this will permit
a determination of the Fermi Constant to \unit[0.5]{ppm}.  I present
an update on our progress and achievements to date.
\end{abstract}

\maketitle

\begin{figure}
{\centering
\begin{minipage}{0.48\textwidth}
\includegraphics[width=\textwidth]{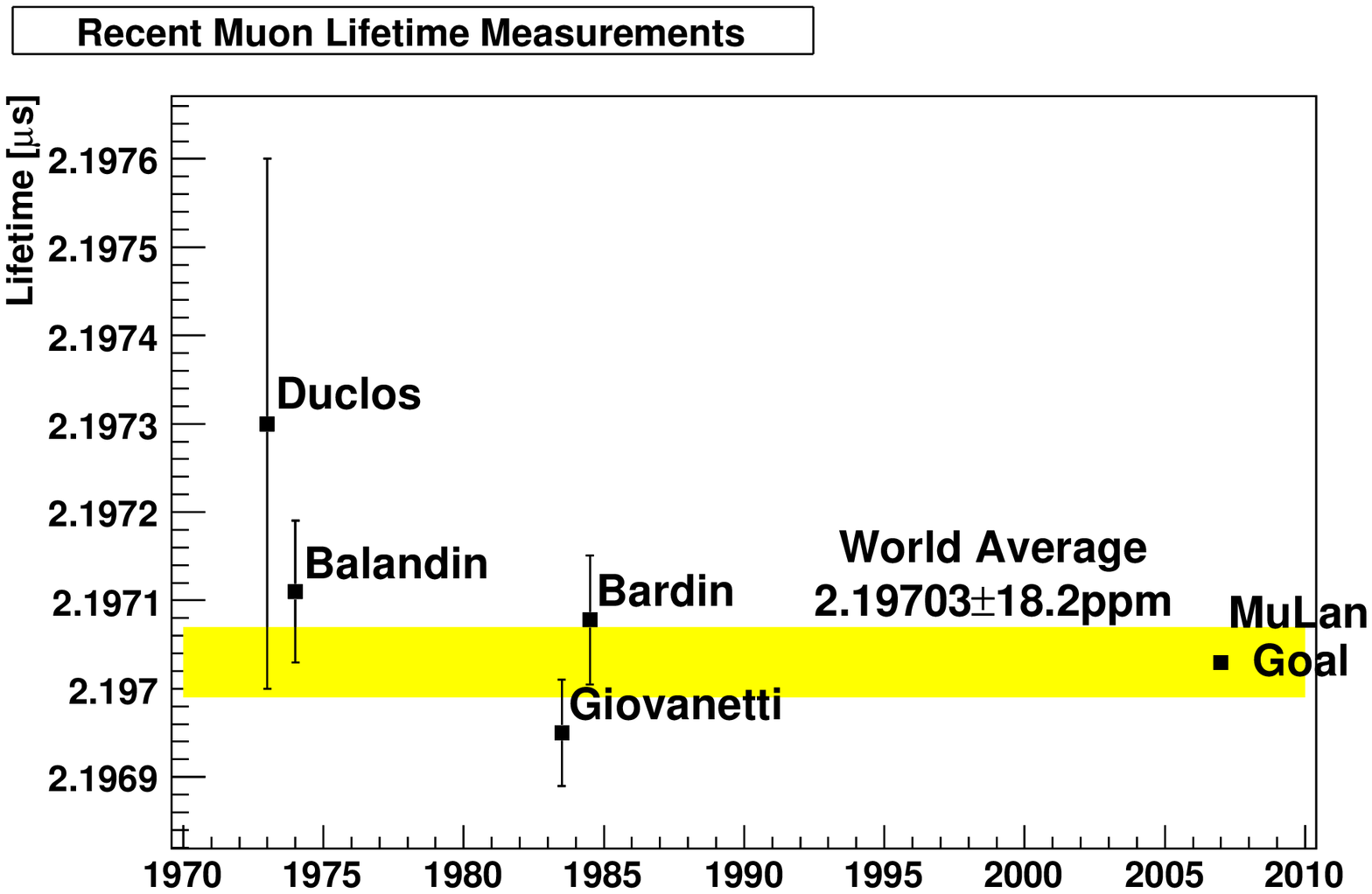}
\end{minipage}
\hfill
\begin{minipage}{0.48\textwidth}
\includegraphics[width=\textwidth]{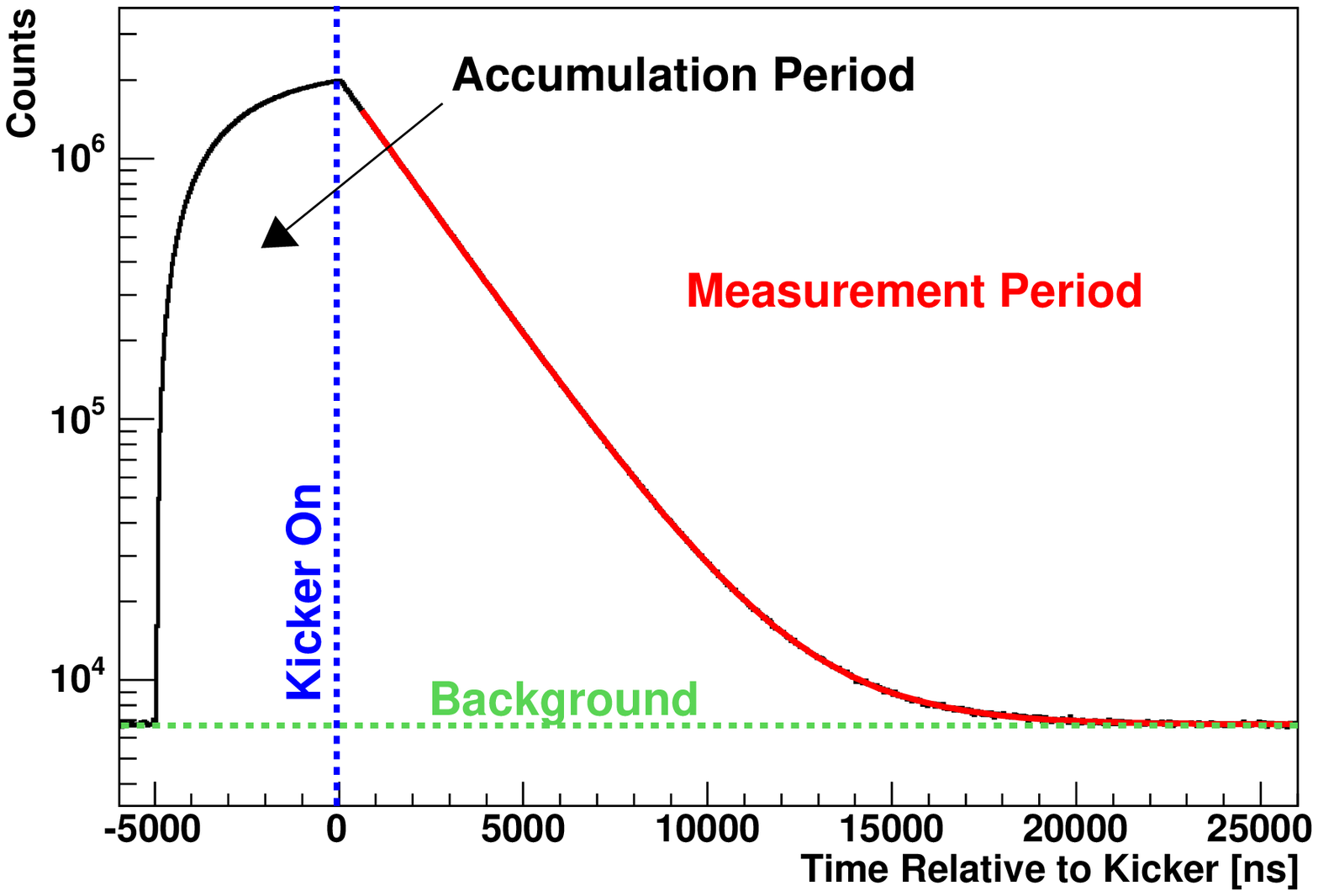}
\end{minipage}
}
\caption{Left: A brief history of muon lifetime
  experiments.\cite{Eidelman:2004wy} The band is the current world
  average of those experiments.  The predicted error bars for the
  MuLan experiment are not visible at this scale. Right: A muon
  lifetime histogram extracted from ten minutes of data collected by
  the MuLan Experiment in 2004.}
\label{figs}
\end{figure}

The Standard Model of electroweak interactions certainly stands as a
triumph of modern physics, due to the impressive agreement among
numerous precision particle and nuclear physics experiments performed
in the last two decades.  During that time, our knowledge of the Fermi
constant, \Gfermi, one of the most fundamental input parameters of the
model, has not changed. The Fermi constant sets the strength of the
weak interaction, and can be cleanly extracted from a measurement of
the free muon lifetime~\cite{vanRitbergen:1999fi}
\begin{gather}
\frac{1}{\tau_\mu} = \frac{\Gfermi^2 m_\mu^5}{192\pi^3} \left(1 +
\sum_i q_i\right)\ .
\label{eq:fermi}
\end{gather}
Here, the $q_i$ encode the non-weak corrections to a tree level
result, including massive phase space ($q_0$) and QED loop corrections
(with $q_1$ the one loop corrections), while all weak interaction
effects are subsumed in \Gfermi.  Until recently, extraction of the
Fermi constant via this relation was theoretically limited by a
\unit[30]{ppm} uncertainty.  A calculation of the two loop QED
corrections ($q_2$ in Equation~\ref{eq:fermi}) by van Ritbergen and
Stuart~\cite{vanRitbergen:1998yd} has reduced the theoretical
uncertainty to the sub-\unit{ppm} level.  Extracting \Gfermi\ from
muon lifetime data is now a statistics limited operation, as no
experiments have been performed in the last two decades (see
Figure~\ref{figs}).

It is not only technically feasible, therefore, but desirable to
improve the experimental situation in both fields with a measurement
of the muon lifetime at the \unit{ppm} level.  This is the charter of
the Muon Lifetime Analysis, or MuLan, experiment~\cite{carey}, an
ongoing effort at the Paul Scherrer Institut (PSI) in Switzerland.
The Collaboration intends to record in excess of $10^{12}$ muon
decays, with systematics controlled to better than the statistical
uncertainty, with the goal of extracting \Gfermi\ to \unit[0.5]{ppm}. 

While previous experiments have utilized low rate, DC muon sources,
such approach does not scale to our statistics.  We require a high
rate, time structured muon source that allows us to perform many muon
lifetime measurements simultaneously.  To this end, we have developed
a high rate (\unit[7]{MHz}) beam tune in the \piEthree\ beam line at
PSI, and constructed a fast electrostatic kicker~\cite{barnes} to chop
the beam.  We collect muons on a fixed target for $\unit[5]{\mu s}$
(two lifetimes) and then observe decays of those muons during a
$\unit[22]{\mu s}$ (ten lifetimes) beam off time window in a large
acceptance, high granularity detector.

The experiment has been designed from the ground up to minimize or
eliminate systematic concerns.  In particular, we must
\begin{enumerate}
\item Minimize spin polarization and precession effects: Muon beams
  are inherently polarized, and the momenta of their decay positrons
  are correlated to their spin direction.  If the muon population in
  our stopping target is allowed to remain polarized, slow spin
  precession in ambient fields would result in substantial
  early-to-late effects during the measurement period.  We combat
  these two effects by using a thin target of Arnokrome-3 (AK3), a
  proprietary alloy with a high internal magnetic field, and a solid
  depolarizing sulfur target with an externally imposed magnetic
  field.  In both cases, the field is oriented perpendicular to the
  beam polarization, and rapidly depolarizes the muon population
  during the beam on period.  Regular rotation of the target (and,
  hence, the field orientation) minimizes the effect of any residual
  polarization.  Additionally, the symmetric detector design allows us
  to minimize residual polarization effects in the sum of point-wise
  elements.
\item Reduce pile-up effects: Our detector is a highly segmented,
  symmetric design, with 170 dual layer detector modules built from
  fast plastic scintillator organized into a truncated icosahedral
  (``soccer ball'') superstructure.  The high granularity assures a
  average occupancy of less than 0.1 hits per detector module per
  kicker cycle.  Custom \unit[500]{MHz} waveform digitizers (WFDs)
  will allow pulse-to-pulse time resolution of better than
  \unit[5]{ns}.  Together, these reduce pile-up corrections to better
  than the $10^{-4}$ level.
\item Beam off muon arrivals: Muons arriving during the beam off
  period reduce the signal to background ratio of our data sample.  To
  minimize the number of out-of-time arrivals, we have designed our
  kicker to operate with a \unit[25]{kV} potential between the plates,
  providing a beam extinction in excess of 1000.  To eliminate any
  early-to-late effects on the background at the \unit{ppm} level from
  a changing extinction factor, the kicker voltage is regulated at
  better than the $10^{-4}$ level for the duration of the measurement
  period.
\item Minimize non-target muon stops: Muons which stop outside the
  high magnetic field of the target are a potential source of
  residually polarized decays.  Even in the absence of polarization,
  these decay do not have uniform acceptance.  To minimize these
  effects, we have constructed a low mass corridor between the end of
  the beam pipe and the target, consisting of thin beam windows and a
  thin Mylar bag between the end of the beam pipe and the target,
  filled with helium.  This construction reduces non-target stops to
  the $10^{-3}$ level.  In the future, we plan to improve the
  situation further, bringing the muons to the target completely in
  vacuum by extending the beam pipe entirely through the detector.
\end{enumerate}

The full MuLan detector was commissioned during a 2003 engineering
run.  In 2004, the kicker was installed in the \piEthree\ beamline,
and we took our first physics data, collecting $10^{10}$ muon decays.
We expect to complete a full analysis of this data before the end of
2006.  The statistical significance will be roughly \unit[8]{ppm},
with comparable systematics.  A ten minute subset of our lifetime data
from this 2004 run is shown in Figure~\ref{figs}.  In 2005 the
WFDs were installed and an engineering run validated the full
experimental setup, collecting almost $10^{11}$ muon decays.  We are
now in the final stages of preparation for a ten week 2006 physics
run, during which we expect to record $10^{12}$ muon decays, with a
statistical reach of \unit[1-2]{ppm}.

\begin{theacknowledgments}
Thanks go to my colleagues in the MuLan Collaboration:
T.I.\,Banks, K.M.\,Crowe, F.E.\,Gray, and B.\,Lauss (University of California,
Berkeley and LBNL);
R.M.\,Carey (Co-spokesman), W.\,Earle, A.\,Gafarov, Z.\,Hartwig,
E.\,Hazen, I.\,Logashenko, K.R.\,Lynch, J.P.\,Miller, Q.\,Peng,
B.L.\,Roberts, W.\,Vreeland, (Boston University);
D.\,Chitwood, S.M.\,Clayton, P.T.\,Debevec, D.W.\,Hertzog
(Co-spokesman), P.\,Kammel, B.\,Kiburg, S.\,Knaack, J.\,Kunkle,
R.\,McNabb, F.\,Mulhauser, D.\,Webber, P.\,Winter (University of
Illinois, Urbana-Champaign); 
C.S.\,\"{O}zben (Istanbul Technical University);
E.\,Bartel, C.\,Church, M.L.\,Dantuono, R.\,Esmaili, K.\,Giovanetti,
M.\,Miller (James Madison University); 
S.\,Battu, S.\,Dhamija, T.\,Gorringe, M.\,Ojha, S.\,Rath,
V.\,Tishchenko (University of Kentucky);
C.J.G.\,Onderwater (Kernfysisch Versneller Instituut, Groningen)
\end{theacknowledgments}

\bibliographystyle{aipproc}

\bibliography{lynch-mulan}

\end{document}